\documentclass[twocolumn,aps,showpacs,showkeys,prb,superscriptaddress,reprint]{revtex4-1}
\usepackage{epsfig}
\usepackage{graphicx}
\usepackage{amsmath}
\usepackage{amssymb}
\usepackage{amsfonts}
\usepackage{tabularx}
\usepackage{color}
\usepackage{dcolumn}
\usepackage{bm}
\usepackage{multirow}

\begin{document}

\title{Parameter-free extraction of EMCD}

\author{S. Muto}
\author{K. Tatsumi}
\affiliation{Department of Materials, Physics, and Energy Engineering, Nagoya University, Chikusa, Nagoya 464-8603, Japan}
\author{J. Rusz}
\affiliation{Department of Physics and Astronomy, Uppsala University, Box 516, S-751\,20 Uppsala, Sweden}
\affiliation{Institute of Physics, Czech Academy of Sciences, Na Slovance 2, CZ-182 21 Prague, Czech Republic}

\begin{abstract}
We present a parameter-free method of extraction of the electron magnetic circular dichroism spectra from energy-filtered diffraction patterns measured on crystalline specimen. The method is based on multivariate curve resolution technique. The main advantage of the proposed method is that it allows extraction of the magnetic signal regardless of the symmetry and orientation of the crystal. In other words, our new method overcomes difficulties caused by complexity of dynamical diffraction effects. 
\end{abstract}


\maketitle

\section{Introduction}

Electron magnetic circular dichroism \cite{nature} (EMCD) is the name for a transmission electron microscopy analogue of the x-ray magnetic circular dichroism (XMCD)\cite{schutz,thole,carra,chen}. Both techniques allow studying magnetic properties of samples via an excitation of a core electron into unoccupied valence states. Due to sharp atom-specific energy levels of the core electrons, this makes both EMCD and XMCD element-selective. Transmission electron microscope (TEM) brings an additional advantage of a superior lateral resolution, which reaches in principle sub-\AA{}ngstr\"{o}m range.

Despite an intense initial activity in EMCD, which led to improvements of spatial resolution \cite{toulouse, lacbed, emcd2nm}, advances in theoretical description \cite{prbtheory, snr, oursr, lionelsr, plasmon, opmaps}, benchmarking quantitative measurements \cite{lionelsr, hansprl, cobalt, warrot} and first applications \cite{nanostuff, klie, bacteria, nanozno}, the experiments are still primarily in the method development stage. The main reasons are high demands on quality of sample preparation, low signal to noise ratio (SNR) and complications due to dynamical diffraction effects.

While the first problem can only be approached from the experimental side, the other two can be aided from the theory side. So far the best control over signal to noise ratio one can obtain by measuring energy-filtered TEM datacubes, using the telefocus mode \cite{lsfollow}. This way one has a two-dimensional set of spectra, which can be processed independently one-by-one to provide a rich statistics for error control. It seems to be generally accepted that the most suitable experimental geometry is the 3-beam orientation, in which one can use the information from all four quadrants in the diffraction plane by means of the double-difference (DD) procedure \cite{hansprl, lsfollow, warrot}.

Here we present a method, how to overcome the requirement of the 3-beam orientation and allow to extract the EMCD signal from arbitrary crystal orientation. The method is based on advanced statistical techniques, namely the multivariate curve resolution (MCR) \cite{mcrals}. Introduction of this method into the EMCD field means that the requirement of a particular crystal orientation is not anymore required -- thus essentially removing the obstacles caused by dynamical diffraction effects.

In the following text we will describe the principles and assumptions of the MCR method (Sec.~\ref{sec:mcr}), method of construction of theoretical datacubes used in this article (Sec.~\ref{sec:dc}) and the results obtained (Sec.~\ref{sec:results}).

\section{Multivariate curve resolution\label{sec:mcr}}

The present multivariate analysis is based on the MCR-alternating least-square (ALS) algorithm\cite{mcrals}, whereby we adopted a faster, accurate and more robust algorithm, called the non-negative matrix factorization (NMF)\cite{paatero,nnmf}. The main improvement of the present NMF is to include smoothness constraints, which are often enforced to regularize the computed solutions in the presence of noise in the data. 

The single EEL spectrum extracted from the datacube at a position corresponding to a single pixel is supposed to be a linear combination of two kinds of spectral components, each consisting of nonmagnetic and either clockwise or counter clockwise chiral components with different weights, depending on the position in reciprocal space. The objective of MCR-ALS is to extract pure component spectra with as few assumptions about the data as possible by repeatedly applying the least-square fit until the residual is converged within the experimental noise level. 

As reported previously \cite{muto09,muto10}, the MCR technique is beset with two inherent difficulties: one is that the number of components underlying cannot be determined a priori; the other is that the result obtained by means of the MCR technique is not unique. The first problem disappears in the present EMCD data. The second one is usually compensated for by including auxiliary constraints on the solutions. We found that the additional smoothness constraint dramatically improved our second problem. The MCR solutions were computed using the NMF method as described in \cite{nnmf}.

\section{Construction of datacubes\label{sec:dc}}

We have tested the MCR technique on theoretical datacubes. The advantage of this approach is that we know precisely the spectral shape of the EMCD signal and its distribution throughout the diffraction plane.

As a benchmark system we chose a bcc iron crystal. The electronic structure was calculated in density functional theory with local density approximation using the code WIEN2k \cite{wien2k} with over 100 basis functions per atom and 40000 k-points in the 1st Brillouin zone, which ensures well converged bandstructure. Based on that, we calculated the MDFFs\cite{kohlrose} using the $\lambda=1$ approximation \cite{prbtheory}. The dynamical diffraction effects were treated using the Bloch's waves formalism described in the same article. We used the automatic Bloch-wave summation technique MATS \cite{mats} with criterion $10^{-5}$, which leads to highly converged calculations in a computationally efficient way.

We have calculated a datacubes in two orientations: 1) A slightly mis-oriented 3-beam orientation so that the extraction of the EMCD signal would be non-trivial,\cite{note1} yet favorable for the DD procedure. The chosen two excited Bragg spots were $\mathbf{G}=\pm(200)$ and Laue circle center $(-0.095, 0.078)G$. 2) A two-beam orientation with an excited Bragg spot $\mathbf{G}=(110)$ and Laue circle center $(\frac{1}{2}\frac{1}{2}0)$. Here we are rather far from the symmetrical 3-beam orientation and we can expect difficulties in EMCD extraction due to asymmetry\cite{2bcasymm}.

In both cases, the energy-filtered diffraction patterns span an area from $-1.2G$ to $1.2G$ along the systematic row and from $-0.7G$ to $0.7G$ in the perpendicular direction, with step $0.025G$ in both directions. The energy range for calculation of MDFFs was from zero up to 15~eV above the Fermi level with step 0.1~eV, separately for both $L_3$ and $L_2$ edges. That means a starting dataset of dimensions $97 \times 57 \times 151$ per edge and thickness. Subsequently, the $L_3$ and $L_2$ edge spectra were first shifted in energy up to 708~eV and 721~eV, respectively, which corresponds to experimental values of onset of the $L_{2,3}$-edges in iron. The spectra were then smeared by 0.7~eV and 1.2~eV wide Lorentzian for $L_3$ and $L_2$ edge, respectively, to account for the finite lifetime of the excitations. Finally, they were overlapped into a single datacube containing both edges. The resulting datacube has an energy range from 695~eV to 745~eV with a step of 0.05~eV. Sample thicknesses were chosen as 10~nm, 20~nm, 30~nm and 40~nm.

On top of the $L_{2,3}$ signals, we added a double-step background signal modeled as a sum of two step functions positioned at 708~eV and 721~eV, respectively, broadened by the same Lorentzians as the edge signal, i.e., 0.7~eV and 1.2~eV, respectively. The relative ratio of heights of the two step functions was set to 2:1. The relative intensity of the post-edge background signal was fixed to 1/3 of the intensity of the broadened nonmagnetic part of the $L_3$ edge peak. These choices reflect the following assumptions: 1) the dynamical diffraction effects have the same influence on the double-step background signal intensity as on the $2p \to 3d$ transitions, 2) the double-step background signal originates from transitions to non-polarized states, and, 3) the life-time of the transitions in question is the same as for $L_{2,3}$ edges. 

Small deviations from these assumptions are possible in experimental situations, for example in \cite{obrien} the polarization of the $s$ states is discussed, with its influence on the extracted orbital to spin moment ratio from XMCD. However, our main aim is to test an accuracy and robustness of the MCR, therefore a study of influence of such effects goes beyond the scope of present manuscript.

Additionally, for each of the four thicknesses, we have also generated datacubes with random noise. Two different noise levels were generated. In particular, we have multiplied the datacubes by a constant factor, so that the $L_3$ peak intensity $I_\mathrm{max}$ becomes 1000 or 10000 counts, respectively, and all the other values were scaled accordingly. Afterwards, at all pixels and energies a random noise value from range $\pm \sqrt{I(q_x,q_y,E)}$ was added. This allowed us to compare the performance of MCR  with DD procedure at various noise levels.

\section{Results\label{sec:results}}

Here we summarize and discuss our results. First we show the raw magnetic signal spectrum and its distribution in the near-3-beam orientation--as a reference for testing the quality of DD and MCR extraction. The datacubes are then analyzed by DD procedure and in the next subsection by MCR. In the last subsection we present an analysis of the 2-beam orientation datasets in a more concise form.

\subsection{Net EMCD signal\label{sec:netemcd}}

\begin{figure}[tbh]
  \includegraphics[width=8.5cm]{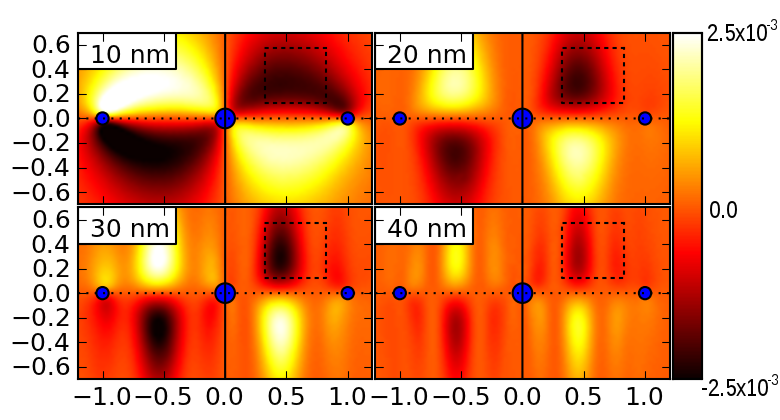}
  \caption{Distribution of the $L_3$-edge integrals of EMCD signal of bcc iron crystal in the diffraction plane at 4 different thicknesses. The energy integral of the $L_3$ edge signal at the transmitted beam is normalized to one. The area marked in the 1st quadrant by a dashed line was used in DD method for extraction of EMCD signal.\label{fig:netemcdimg}}
\end{figure}

\begin{figure}[tbh]
  \includegraphics[width=8.5cm]{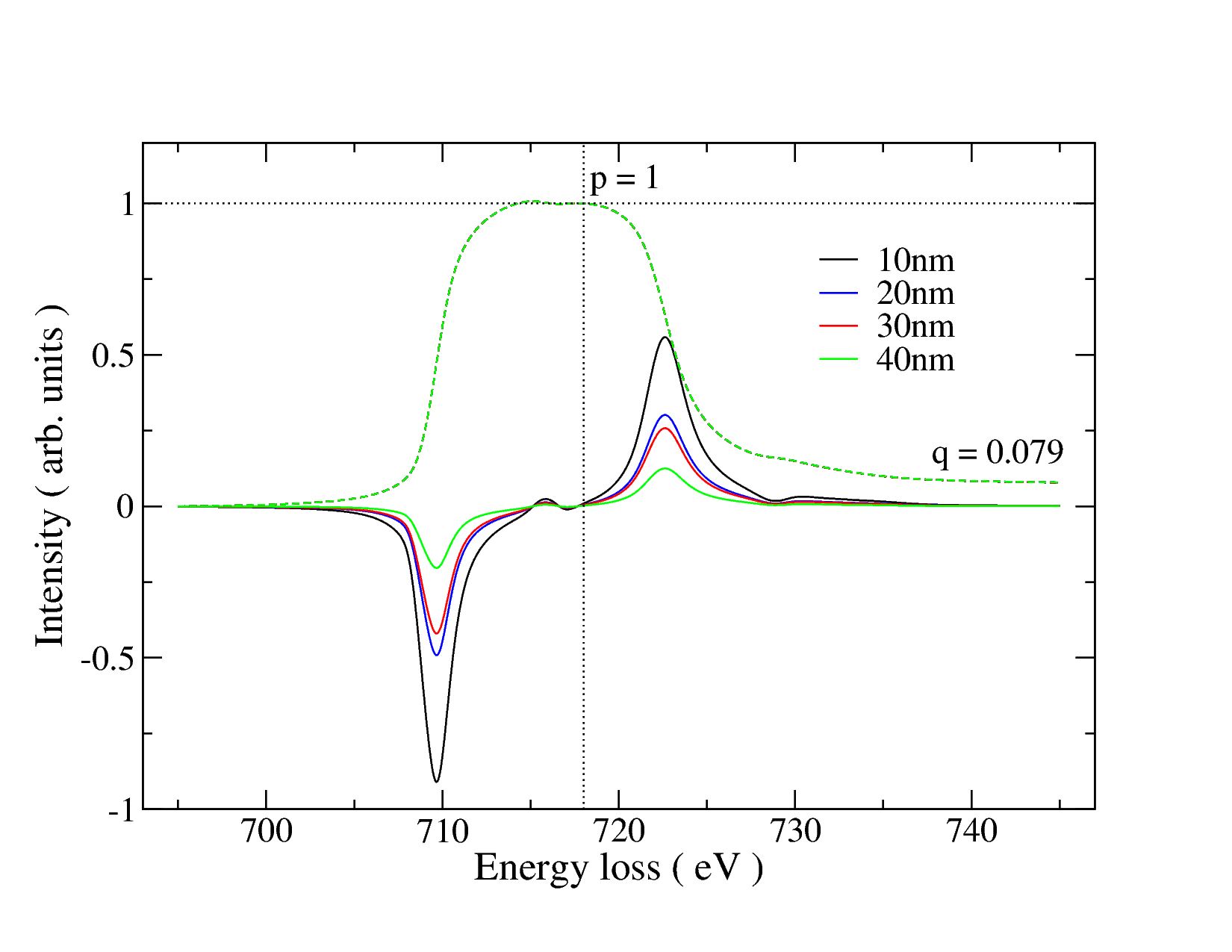}
  \caption{Averaged EMCD spectra (full lines) and their energy integrals (dashed lines) of bcc iron crystal in the diffraction plane at 4 different thicknesses. The energy-integrated EMCD spectrum is scaled so that its value $p$ at 718~eV equals to 1. Post-edge value $q$ enters the sum-rule expressions, see text for details.\label{fig:netemcdspec}}
\end{figure}

As indicated in the previous section, for a tilted 3-beam orientation there is no symmetry operation, that would allow an exact extraction of the EMCD signal from the datacube. The DD procedure was devised to minimize the errors introduced due to misorientation and it was successfully applied in \cite{hansprl, warrot}. Here we work with a theoretical datacube and thus we have access to the EMCD signal by construction. Namely, we calculated the initial dataset in two steps, first only considering the real parts of the MDFF and second, considering only the imaginary parts of the MDFF. It is the latter that contains a net EMCD signal, neglecting the tiny magnetic dipole term \cite{opmaps}. By applying the Lorentzian broadening and overlapping $L_3$ and $L_2$ edges, as discussed above, we obtain datacubes containing net magnetic signal. This serves as a reference dataset, against which we can test the results of the DD and MCR analysis. 

Fig.~\ref{fig:netemcdimg} shows the distribution of the EMCD signal in the diffraction plane. In the figure we show an energy integral of the EMCD at $L_3$ edge and the color bar range is relative to the transmitted beam intensity at $L_3$ edge. We can see that the overal intensity of the EMCD signal decreases with thickness and the pattern of its distribution becomes more complicated due to stronger dynamical diffraction effects at larger thicknesses. The maximum intensity of EMCD signal reaches around 0.3\% of the transmitted beam intensity -- the major reason of difficulties with obtaining high SNR in experiments.

Fig.~\ref{fig:netemcdspec} summarizes net EMCD spectra averaged over the diffraction plane. The averaged spectra were obtained by summing all EMCD spectra, choosing the sign of $L_3$ magnetic contribution as negative. The plot confirms that the magnetic signal decreases with thickness, as anticipated from Fig.~\ref{fig:netemcdimg}. By scaling the individual EMCD curves to the same peak magnitudes, the shapes of spectra become practically identical. This is because the magnetic properties in our model are based on electronic structure of bulk iron regardless of sample thickness considered in the dynamical diffraction calculation.

In quantitative EMCD experiments \cite{hansprl, warrot}, the EMCD spectrum is analyzed by sum rule expressions \cite{oursr, lionelsr} in order to extract the ratio of orbital to spin magnetic moment. We will apply these sum rules as a sensitive test of the accuracy of EMCD spectrum extraction. The sum rules state
\begin{equation}
  \frac{m_l}{m_s} = \frac{2}{3} \frac{\int_{L_3} \!\!\! \Delta\sigma(E) \mathrm{d}E + \int_{L_2} \!\!\! \Delta\sigma(E) \mathrm{d}E}{\int_{L_3} \!\!\! \Delta\sigma(E) \mathrm{d}E - 2\int_{L_2} \!\!\! \Delta\sigma(E) \mathrm{d}E} = \frac{2q}{9p-6q} \label{eq:mlms}
\end{equation}
where $q$ is an energy integral of EMCD over both edges and $p$ is an energy integral over the $L_3$ edge only. Without loss of generality we can rescale the EMCD spectrum or its integral so that $p=1$. Then the $m_l/m_s$ ratio is a function of $q$ only, allowing easy visual comparison of the energy integrals in the post-edge region.

In this manuscript, the integral over $L_3$ edge will mean an energy range from 695~eV to 718~eV.  There is a certain degree of ambiguity in this choice, however we have checked that none of our conclusions do substantially depend on this choice.

Performing the energy integral of the scaled ($p=1$) net EMCD spectra leads to a value of $q=0.079$ which gives $m_l/m_s = 0.019$. This is in good agreement with the DFT values $m_l = 0.048\mu_B$ and $m_s=2.26\mu_B$, i.e., $m_l/m_s = 0.021$. The error can be attributed to energy dependence of the radial wavefunctions in matrix elements \cite{wu}.

\subsection{Double difference procedure\label{sec:ddemcd}}

In Ref.~\cite{hansprl} we have introduced the DD procedure on post-edge normalized datacube. This method efficiently cancels the systematic errors introduced by slight mis-orientations from a perfect 3-beam orientation \cite{lsfollow}.  

When applying DD, we first normalize the datacube in the post-edge region, so that the energy integral over the last  2.5~eV is equal for all spectra in the datacube. (We will discuss later the dependence of results on chosen normalization range.) Then for every energy-loss, a mirror image of the lower diffraction half-plane is subtracted from the upper diffraction half-plane (1st difference) and the lower half-plane is removed from further considerations. Finally, we subtract a mirror image of the left quadrant from the right quadrant (2nd difference) and we remove the left quadrant from further considerations. I.e., the dataset is reduced by DD to one quarter of its original size. Now a spectrum at every pixel corresponds to an approximation of the EMCD spectrum, assuming an efficient cancellation of non-magnetic components by the DD. An averaged EMCD spectrum can be constructed either from the whole quadrant or some selected region, as was done in \cite{hansprl}.

We will apply this method of extraction of EMCD spectra to our theoretical datacubes (both noise-less and noisy ones) to a) test the accuracy of the DD procedure on realistic model data with known magnetic component and b) to compare its performance to the new method of extraction of EMCD based on MCR discussed in the next subsection.

First we have extracted averaged EMCD spectra from the whole quadrant. At selected four sample thicknesses, from the noise-less datacube we obtained stable $m_l/m_s$ value of $0.029$. For the noisy datacube with peak-value $I_\mathrm{max}=10^4$ the span of $m_l/m_s$ values was $0.023$--$0.028$ and for the most noisy datacube ($I_\mathrm{max}=10^3$) the span was from $-0.013$ to $0.052$, which is a large error.

\begin{figure}[tbh]
  \includegraphics[width=8.5cm]{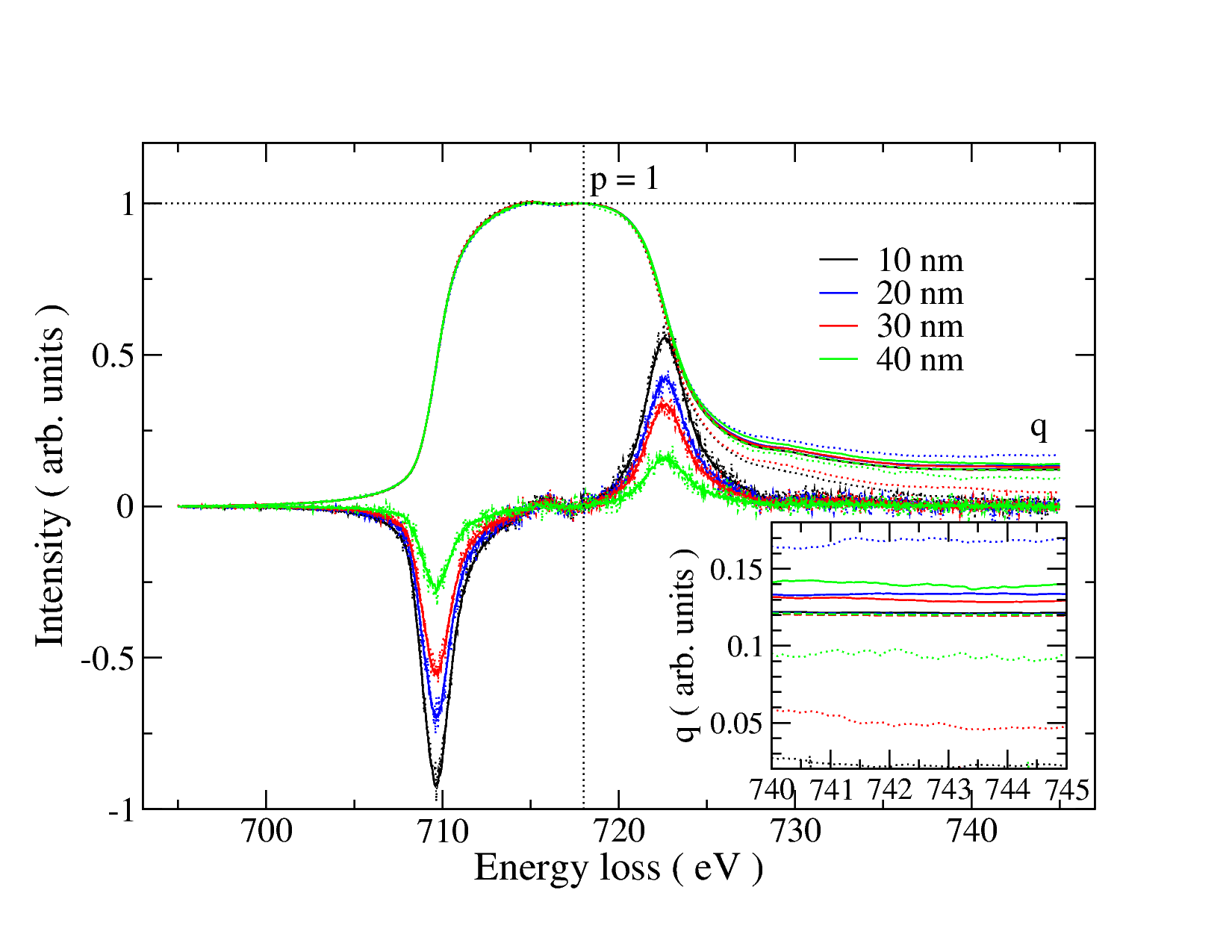}
  \caption{EMCD spectra and their energy integrals for bcc iron crystal at 4 different thicknesses obtained by the double-difference procedure from a region marked in Fig.~\ref{fig:netemcdimg}. The energy-integrated EMCD spectrum is scaled so that its value $p$ at 718~eV equals to 1. The inset zooms in on the post-edge region of the energy integrals, $q$. The values of $q$ enter the sum-rule expressions for evaluation of $m_l/m_s$ ratio, see text for details. Full, dashed and dotted lines correspond noise-less, $I_\mathrm{max}=10^4$ and $10^3$, respectively.\label{fig:ddemcdspec}}
\end{figure}

While the result is quite satisfactory for noise-less or low-noise ($I_\mathrm{max}=10^4$) datacubes, for the most noisy datacube the accuracy of $m_l/m_s$ extraction is quite poor. The reason is that in DD the non-magnetic part of the spectra is almost removed and in areas of zero or low magnetic signal we are adding basically only noise and eventual anisotropy signal. That was also the reason for using a limited integration window around an area of the strongest magnetic signal instead of the whole diffraction plane in \cite{hansprl}. That introduces some additional parameters, such as position, shape and size of the integration window. Here we have used a rectangular region around the so-called Thales circle position \cite{nature} with dimensions $0.5G \times 0.45G$, similarly as in \cite{hansprl,lsfollow}, see Fig.~\ref{fig:netemcdimg}. We have tested that small variations in the position and size of the region do not qualitatively change the results.

Use of the limited integration window led again to stable results for the noise-less data: we obtained $q=0.121 \pm 0.003$, which gives $m_l/m_s = 0.029$. Note that this value over-estimates the expected value $0.019$ based on raw EMCD evaluation in Sec.~\ref{sec:netemcd}. Our analysis shows that this offset is a systematic error and that it comes from the post-edge normalization procedure itself, see Appendix~\ref{sec:normalization}. The error is approximately $0.01$ for our dataset and is almost independent of the value of $m_l/m_s$ ratio itself. The magnitude of the error depends on the chosen broadening model and range of energy-integrals over edges, particularly in the post-edge region.

With increasing noise in datacubes the accuracy of DD relatively quickly drops down, see Fig.~\ref{fig:ddemcdspec}. For the datacube with peak-value $10^4$ the DD procedure provided values $m_l/m_s = 0.029$--$0.035$. At highest noise level -- datacube with peak-value 1000 -- the DD procedure provided large spread of values $m_l/m_s = 0.007$--$0.044$ (note the order of magnitude spread), yet somewhat improved in comparison to the evaluation based on the whole quadrant.

\subsection{MCR method\label{sec:mcremcd}}

Results obtained by MCR method are described here. First we discuss the efficiency of decomposition of the spectra in terms of maps of coefficients and the residuals and then we summarize the extracted $m_l/m_s$ ratios. Note that in MCR method we do not use any integration window as in DD procedure. The reason is that MCR efficiently uses all data available to strengthen the model of both magnetic and non-magnetic spectral components, i.e., areas with low EMCD signal help to improve the model for non-magnetic spectral component, which in turn effectively improves the accuracy of extraction of the magnetic one in areas with stronger EMCD signal. In this sense, use of a limited integration window can be counter-productive for MCR and tests performed on our datasets have confirmed that (not shown).

\begin{figure}[tbh]
  \includegraphics[width=8.5cm]{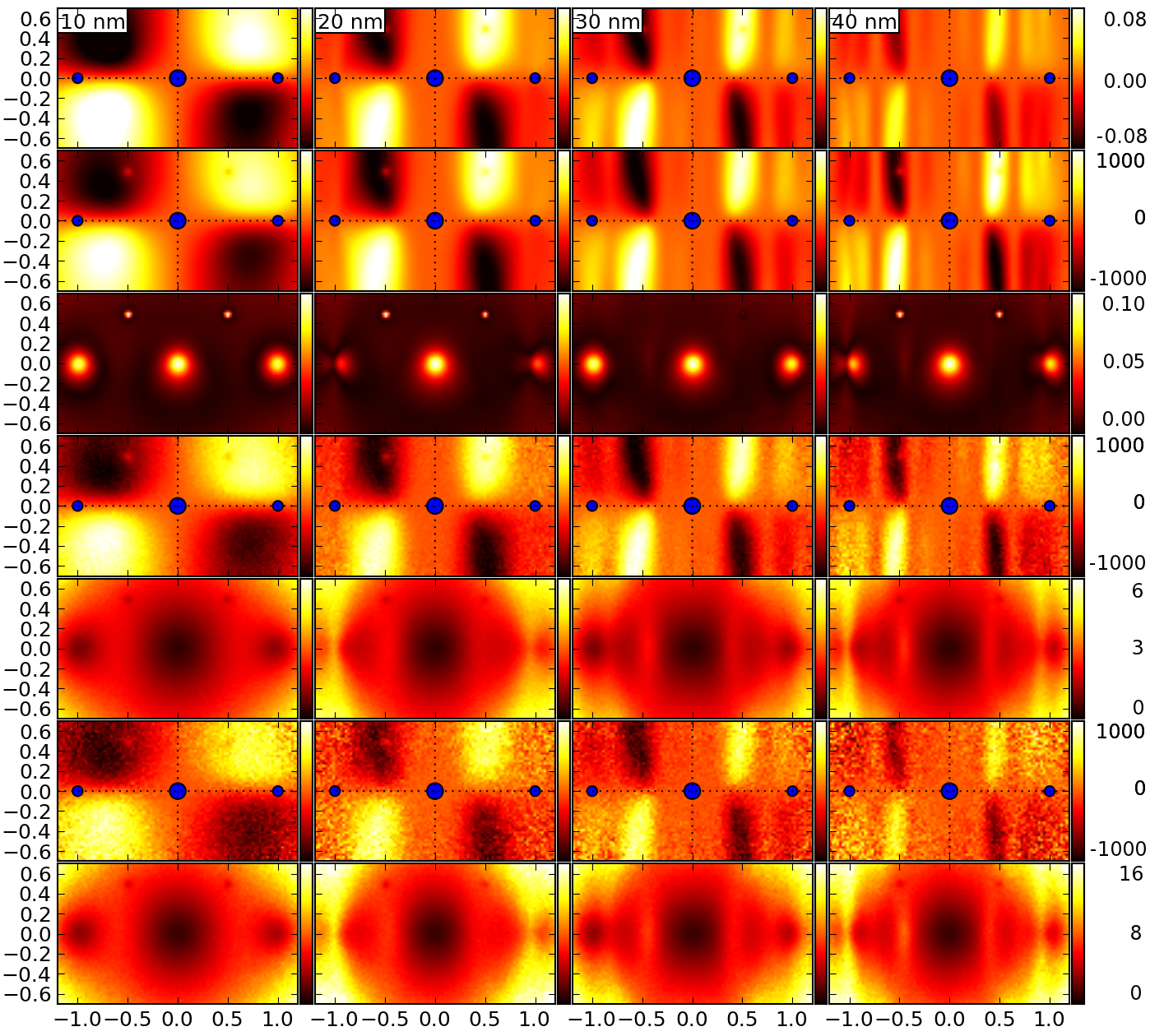}
  \caption{Maps of distributions of EMCD signal and residui of MCR fits. Row 1 shows distribution of the raw magnetic signal in post-edge normalized datacube (cf.\ Fig.~\ref{fig:netemcdimg}). Row 2 shows map of coefficients of the EMCD signal extracted by MCR on noise-free datacube. Row 3 shows an associated map of residui. Rows 4--5, and 6--7 show the same as rows 2--3, but for the noisy datacube with peak level $10^4$ and $10^3$, respectively. Each column corresponds to different thickness, as indicated in the top row.\label{fig:mcrmaps}}
\end{figure}

The MCR procedure outlined in Sec.~\ref{sec:mcr} produced stable fits for all 12 tested datacubes. Every such fit results in isolation of two spectral components along with maps of corresponding positive coefficients throughout the diffraction plane. The distribution of the EMCD signal is then obtained as a difference of coefficients of the two spectral components. These maps are summarized in rows 2, 4 and 6 of Fig.~\ref{fig:mcrmaps} for noise-free, $I_\mathrm{max}=10^4$ and $I_\mathrm{max}=10^3$ datacubes, respectively. These maps can be compared to theoretical relative dichroic maps shown in the top row and an excellent agreement is found. The only differences are due to noise, but the position and shape of regions of positive or negative magnetic signal is recovered with high accuracy.

A more quantitative measure of the quality of the fit is expressed in terms of maps of residui. The value of a residuum at a particular pixel $(q_x,q_y)$ is defined as $\sqrt{\sum_{i} [\Delta I(q_x,q_y,E_i)]^2}$, where $\Delta I(q_x,q_y,E_i)$ represents the difference between the actual and fitted value of the pixel at energy $E_i$. For the noise-free datacube the map of residui shows negligible values demonstrating that the model fitted by MCR represents data with high accuracy. Small deviations are found only at positions of Bragg spots, but even there the error is negligible (see the color scale).

For the noisy datacubes, residues necessarily can't be zero. Overall, SNR is approximately 3.2 times higher for the datacube with $I_\mathrm{max}=10^4$ compared to the datacube with $I_\mathrm{max}=10^3$. That can be explained by the nature of Poissonian noise: the datacube with $I_\mathrm{max}=10^4$ has 10 times stronger signal, the noise increases by factor of $\sqrt{10}$ and in consequence SNR improves by factor of $10/\sqrt{10} \approx 3.2$ -- in agreement with the scales of the maps of residui.

The shape of the noise distribution can be explained via the normalization. Since the intensity of the signal in the corners is weaker than around the transmitted beam or Bragg spots, SNR is lower in the corners. After the post-edge normalization of spectra, the noise in low-signal regions becomes amplified.

It is interesting to note, that noise is the dominant source of residui in our noisy datacubes -- making the fit more accurate around peak positions, while it is less accurate in low signal areas. For the noise-free datacube it is the other way around, see row 3 in Fig.~\ref{fig:mcrmaps} and compare with rows 5 and 7. It indicates that the spectral shape is slightly different at Bragg spots, compared to regions in between. Since the Bragg spots form only a fraction of the area of diffraction plane, in presence of zero noise they show the highest residual signal. It would be interesting to understand the nature of this very weak, but systematic residual signal, however that goes beyond the scope of the present study.

\begin{figure}[tbh]
  \includegraphics[width=8.5cm]{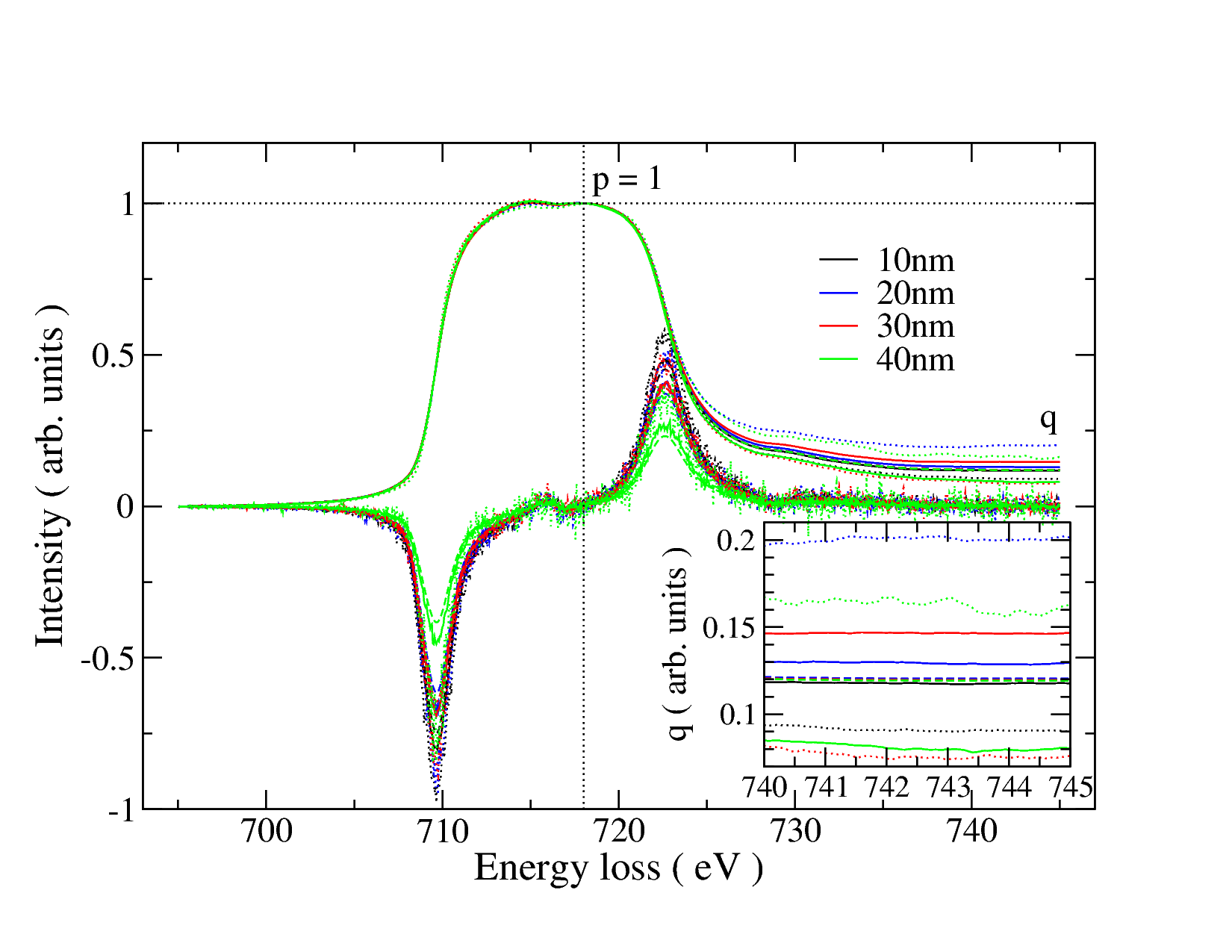}
  \caption{EMCD spectra and their energy integrals of bcc iron crystal at 4 different thicknesses obtained by the MCR method. See caption of Fig.~\ref{fig:ddemcdspec} and text for details.\label{fig:mcremcdspec}}
\end{figure}

Fig.~\ref{fig:mcremcdspec} summarizes the quantitative analysis of the EMCD spectra extracted by MCR method. For the noise-less datacube we obtain in agreement with DD procedure a value of $q=0.120$ with a spread of less than 0.001, leading to a value $m_l/m_s=0.029$ -- again containing the systematic error originating from the post-edge normalization, Appendix~\ref{sec:normalization}. For the noisy datacube with $I_\mathrm{max}=10^4$ we obtained $m_l/m_s$ ratio ranging from 0.019 to 0.036. This spread is comparable to the DD procedure, however, contrary to DD procedure, in MCR the range of values is centered around $m_l/m_s = 0.029$ extracted from noise-less datacube. Finally, for the noisy datacube with $I_\mathrm{max}=10^3$ we obtained $m_l/m_s$ values ranging from 0.017 to 0.051.

This shows that in a near-3-beam orientation, which is optimal for application of the DD procedure, the MCR method is at least as efficient in extraction of EMCD spectra as the DD procedure. In the next subsection we will compare performance of both methods far from 3-beam orientation, where DD is less applicable.

\begin{table}[thb]
\caption{Summary of $m_l/m_s$ extraction in near-three-beam orientation. Net values are obtained directly from averaged EMCD spectra. Then DD and MCR methods were applied on noise-free and noisy datacubes. The level of noise is expressed via the maximum intensity $I_\mathrm{max}$ at the $L_3$ peak.\label{tab:3bcdata}}
\begin{tabular*}{8.6cm}{@{\extracolsep{\fill}}lcccc}
\hline\hline
  & Net & $I_\mathrm{max}$ & DD & MCR \\
\hline
\multirow{3}{*}{10 nm} & \multirow{3}{*}{0.0184} & $\infty$ & $0.029 \pm 0.002$ & $0.029 \pm 0.002$ \\
                       &                         &  $10^4$  & $0.030 \pm 0.002$ & $0.027 \pm 0.003$ \\
                       &                         &  $10^3$  & $0.010 \pm 0.009$ & $0.024 \pm 0.006$ \\
\hline
\multirow{3}{*}{20 nm} & \multirow{3}{*}{0.0185} & $\infty$ & $0.029 \pm 0.002$ & $0.030 \pm 0.002$ \\
                       &                         &  $10^4$  & $0.034 \pm 0.002$ & $0.028 \pm 0.006$ \\
                       &                         &  $10^3$  & $0.040 \pm 0.010$ & $0.043 \pm 0.013$ \\
\hline
\multirow{3}{*}{30 nm} & \multirow{3}{*}{0.0184} & $\infty$ & $0.029 \pm 0.002$ & $0.029 \pm 0.002$ \\
                       &                         &  $10^4$  & $0.029 \pm 0.008$ & $0.035 \pm 0.003$ \\
                       &                         &  $10^3$  & $0.017 \pm 0.017$ & $0.015 \pm 0.015$ \\
\hline
\multirow{3}{*}{40 nm} & \multirow{3}{*}{0.0185} & $\infty$ & $0.029 \pm 0.002$ & $0.029 \pm 0.002$ \\
                       &                         &  $10^4$  & $0.030 \pm 0.012$ & $0.019 \pm 0.011$ \\
                       &                         &  $10^3$  & $0.033 \pm 0.018$ & $0.033 \pm 0.034$ \\
\hline\hline
\end{tabular*}
\end{table}

In the Table~\ref{tab:3bcdata} we summarize and extend the results of both DD and MCR. We have included an error bar, which comes from an analysis of the obtained $m_l/m_s$ ratio as a function of post-edge normalization interval, namely from 1~eV up to 5~eV. In general, we have observed the following behavior: a) for too narrow intervals, particularly in noisy datasets, the extracted $m_l/m_s$ ratio can vary strongly--therefore we picked 1~eV as the smallest range; b) for too large intervals the obtained $m_l/m_s$ ratio slowly and monotonously grows, because the fraction of the magnetic signal in the post-edge region increases, see Eq.~(\ref{eq:mpe}).

\subsection{Two-beam orientation\label{sec:2bc}}

\begin{figure}[tbh]
  \includegraphics[width=8.5cm]{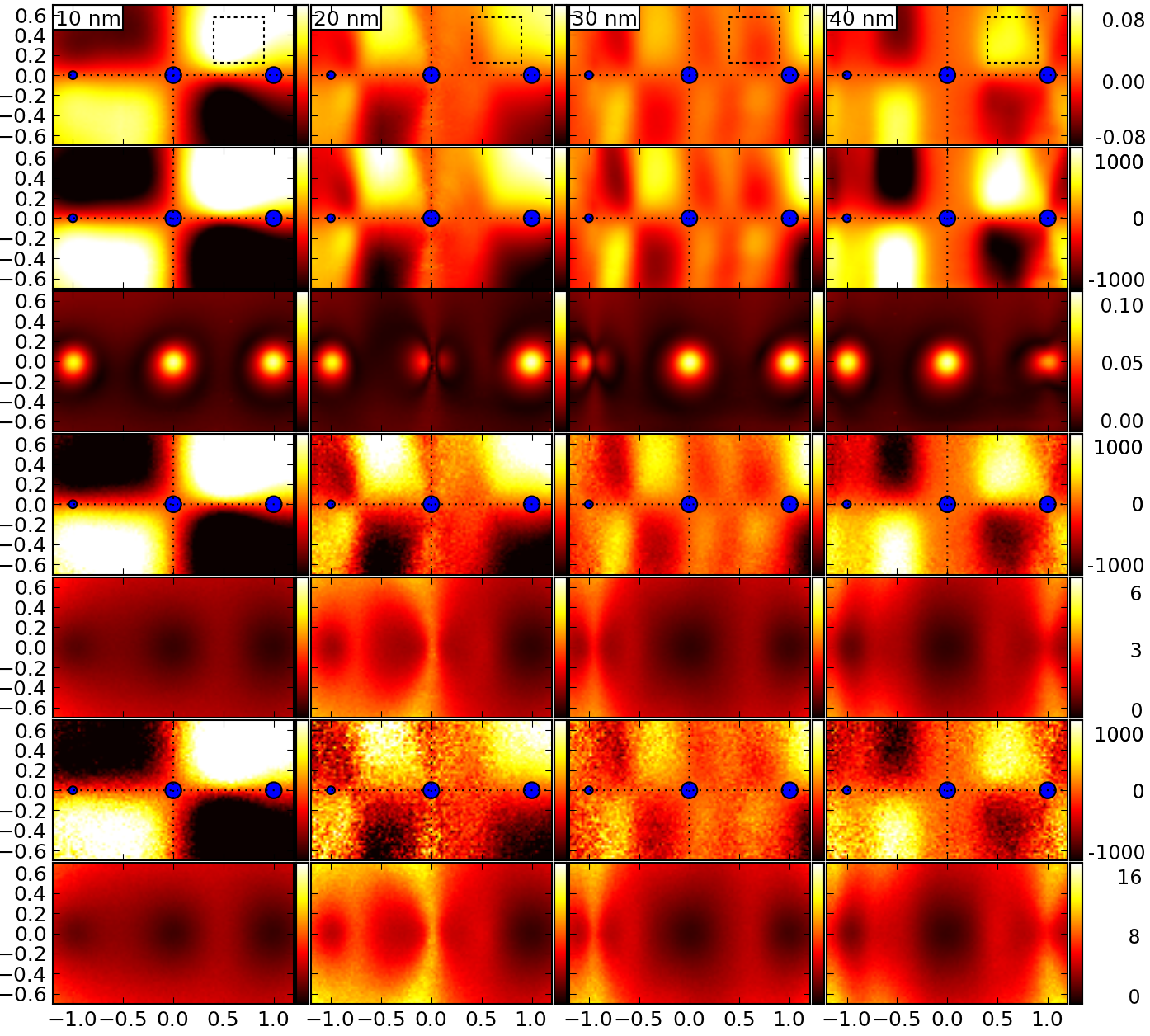}
  \caption{Same as Fig.~\ref{fig:mcrmaps}, but in two-beam orientation with $\mathbf{G}=(110)$. The integration window tested with DD procedure is indicated in the top row.\label{fig:2bc_mcr_maps}}
\end{figure}

\begin{table}[b]
\caption{Summary of $m_l/m_s$ extraction in two-beam orientation. Net values are obtained directly from averaged EMCD spectra. Then DD and MCR methods were applied on noise-free and noisy datacubes. The level of noise is expressed via the maximum intensity $I_\mathrm{max}$ at the $L_3$ peak.\label{tab:2bcdata}}
\begin{tabular*}{8.6cm}{@{\extracolsep{\fill}}lcccc}
\hline\hline
  & Net & $I_\mathrm{max}$ & DD & MCR \\
\hline
\multirow{3}{*}{10 nm} & \multirow{3}{*}{0.0182} & $\infty$ & $0.029 \pm 0.002$ & $0.029 \pm 0.001$ \\
                       &                         &  $10^4$  & $0.027 \pm 0.003$ & $0.029 \pm 0.002$ \\
                       &                         &  $10^3$  & $0.031 \pm 0.007$ & $0.028 \pm 0.003$ \\
\hline
\multirow{3}{*}{20 nm} & \multirow{3}{*}{0.0184} & $\infty$ & $0.024 \pm 0.002$ & $0.030 \pm 0.001$ \\
                       &                         &  $10^4$  & $0.15 \pm 0.10$ & $0.028 \pm 0.003$ \\
                       &                         &  $10^3$  & $0.34 \pm 0.38$ & $0.031 \pm 0.016$ \\
\hline
\multirow{3}{*}{30 nm} & \multirow{3}{*}{0.0181} & $\infty$ & $0.038 \pm 0.002$ & $0.029 \pm 0.001$ \\
                       &                         &  $10^4$  & $0.03 \pm 0.19$ & $0.032 \pm 0.002$ \\
                       &                         &  $10^3$ & $-0.29 \pm 0.10\phantom{0}$ & $0.070 \pm 0.019$ \\
\hline
\multirow{3}{*}{40 nm} & \multirow{3}{*}{0.0180} & $\infty$ & $0.029 \pm 0.002$ & $0.029 \pm 0.001$ \\
                       &                         &  $10^4$  & $0.023 \pm 0.005$ & $0.022 \pm 0.004$ \\
                       &                         &  $10^3$  & $0.032 \pm 0.014$ & $0.025 \pm 0.009$ \\
\hline\hline
\end{tabular*}
\end{table}

Raw EMCD spectra in the two-beam orientation are expectedly very similar to those in Fig.~\ref{fig:netemcdspec}, providing also the same $m_l/m_s$ ratio. The DD procedure did not perform very well, especially for 20~nm and 30~nm datasets, see Table~\ref{tab:2bcdata}. The reason of poor results at these thicknesses can be understood due to low relative strength of the magnetic signal and also due to varying sign of the magnetic signal within a quadrant, see Fig.~\ref{fig:2bc_mcr_maps}. By fine-tuning the size and shape of the virtual aperture it might be possible to improve the results of DD procedure. On the other hand, this shows the potential caveats of this method. Note that in Ref.~\cite{hansprl} only an up-down difference was applied to extract the EMCD signal in the two-beam orientation. Since we are far from three-beam orientation, one might wonder whether a simple difference would be more accurate. We have tested this, however the results with simple up-down difference were worse or comparable with DD procedure (not shown).

In the same way as in the near-3-beam orientation, the MCR method demonstrated its robustness also here. The EMCD spectra were easily extracted and the maps of magnetic signal well corresponded with the expected calculated distributions, see Fig.~\ref{fig:2bc_mcr_maps}. The $m_l/m_s$ ratios were extracted with high accuracy, except for the noisiest datacubes at thicknesses 20~nm and 30~nm, see Table~\ref{tab:2bcdata}. However, even there the MCR method provided values closer to expected $m_l/m_s=0.028$ than DD procedure.

\section{Conclusions}

We have developed a new method for extraction of the EMCD spectra using multivariate curve resolution (MCR). We have shown that with this method we can extract the EMCD spectrum with the same or higher precision than with the double-difference procedure. The main advantage of MCR method is its generality -- it does not require presence of any symmetry planes. The MCR method is therefore applicable in any geometry, not only in the near-3-beam orientation required by double-difference procedure. In principle, by use of MCR method we avoid the obstacles caused by dynamical electron diffraction. Our results will simplify future analyses of the experimental EFTEM datacubes and extraction of the EMCD spectra.

\section{Acknowledgements}

J.R.\ acknowledges the support of Swedish Research Council and STINT. Part of simulations was performed on computer cluster \textsc{Dorje} at Czech Academy of Sciences. This work is supported in part by a Grant-in-Aid for Scientific Research (KAKENHI) in Priority Area (\#474) ``Atomic Scale Modification'' from MEXT, Japan.

\appendix

\section{Brief description of MCR-ALS method}

The datacube is rearranged into a two dimensional matrix form, each column corresponding to a single spectrum extracted from a single pixel of the datacube. Given $m \times n$ matrix $\mathbf{X}$ representing the datacube, $m \times k$ matrix of concentration profiles $\mathbf{C}$, and $n \times k$ matrix of pure component spectra $\mathbf{S}$, where $k$ is the number of components, the following relation holds
\begin{equation} \label{eq:mcrmodel}
\mathbf{X} = \mathbf{C}\mathbf{S}^\intercal + \mathbf{E},
\end{equation}
where the superscript $\intercal$ denotes the transpose of the matrix. $\mathbf{E}$ is the residual matrix with the data variance or statistical noise unexplained by $\mathbf{C}\mathbf{S}^\intercal$. The goal is to obtain physically meaningful $\mathbf{C}$ and $\mathbf{S}$. The data set consists of a set of intensities $I(x, y, E)$ at the spatial position $(x, y)$ and energy-loss $E$. The data can thus be expressed by the three dimensional array $\mathbf{X}(m_x, m_y, n)$, whose variables respectively correspond to the pixel (channel) coordination numbers of the $x$ and $y$ directions and energy-loss $E$. However, the spatial coordinates can be equivalently treated and thus the data is essentially bilinear, expressed by the array $\mathbf{X}(m, n)$, where $m = m_x \times m_y$. Now each row in the matrix $\mathbf{S}^\intercal$ on the right hand side of Eq.~(\ref{eq:mcrmodel}) stands for a pure component EEL spectrum and if each row vector of $\mathbf{S}^\intercal$ is normalized in its intensity (and the total core-loss cross sections are not very different for the component spectra), the $m$-th value of the $k$-th column of the matrix $\mathbf{C}$ denotes the relative composition of the $k$-th component spectrum ($k$-th row of $\mathbf{S}^\intercal$) at the $m$-th spatial coordination. This is a two-way bilinear model and the alternating least-square (ALS) method is usually utilized as a solution of a straightforward regression problem where initial estimates for $\mathbf{C}$ and $\mathbf{S}^\intercal$ are generated initially using either random numbers, eigenvalue decomposition, or dissimilarity criterion.

\section{Influence of normalization on quantitative results\label{sec:normalization}}

Schematically we can decompose a spectrum into nonmagnetic part $N$, magnetic part $M$ and double-step background signal $B$:
$$
\sigma(E,q_x,q_y) = N(E) + M(E) + B(E)
$$

In the post-edge (PE) region the following relations can be extracted from our datacubes:
\begin{eqnarray}
  B_{PE} &    =    & N_{L_3,\mathrm{max}}/3 \\
  M_{PE} & \approx & -0.002 M_{L_3,\mathrm{max}} \label{eq:mpe} \\
  N_{PE} & \approx & 0.001 N_{L_3,\mathrm{max}}
\end{eqnarray}
These values naturally depend on the chosen broadening model and the definition of the PE area.

Generally the magnetic signal is at most $10$--$15\%$ of the non-magnetic signal at $L_3$, i.e.,
\begin{equation}
  |M_{L_3,\mathrm{max}}| \lesssim 0.15 N_{L_3,\mathrm{max}}, \label{eq:mnmax}
\end{equation}
therefore the post-edge intensity of the spectrum in our model datasets is
\begin{equation}
  \sigma(q_x,q_y)_{PE} \approx 0.001 N_{L_3,\mathrm{max}} + \frac{N_{L_3,\mathrm{max}}}{3} - 0.002 M_{L_3,\mathrm{max}}
\end{equation}
where we used relation $B_{PE}=N_{L_3,\mathrm{max}}/3$ applied in construction of our datacubes.

To simplify the notation, we'll introduce the following
\begin{eqnarray}
  \sigma(q_x,q_y)_{PE} & = & C ( 1 \pm a ) \\
  C & = & N_{L_3,\mathrm{max}} \left( \frac{1}{3} + 0.001 \right) \\
  a & = & 0.002 | M_{\mathrm{max}} | / C \label{eq:amc}
\end{eqnarray}
Using this notation and Eq.~(\ref{eq:mnmax}), the value of 
$$a < 0.002 \times 0.15 N_{L_3,\mathrm{max}} / C \approx 0.0001 \ll 1$$
and therefore
\begin{equation}
 [C(1\pm a)]^{-1} \approx (1\mp a)/C.
\end{equation}
That will be used below in normalization of spectra.

Now we will turn our attention to energy integrated spectra. We need separate energy-integrals over the $L_3$ and $L_2$ edges, respectively. We take a difference of two PE-normalized spectral integrals $\tilde{\sigma}_+,\tilde{\sigma}_-$ to obtain edge-integrals of the approximate magnetic signal $\Delta\tilde{\sigma}$:
\begin{eqnarray}
  \tilde{\sigma}_+ & = & (N_{L_3} + M_{L_3} + B_{L_3} + N_{L_2} - M_{L_2} + B_{L_2})(1+a)/C \nonumber \\
  \tilde{\sigma}_- & = & (N_{L_3} - M_{L_3} + B_{L_3} + N_{L_2} + M_{L_2} + B_{L_2})(1-a)/C \nonumber \\
  \Delta\tilde{\sigma} & = & \frac{2}{C}[ M_{L_3} + a(N_{L_3} + B_{L_3}) - M_{L_2} + a(N_{L_2} + B_{L_2})] \nonumber \label{eq:deltasig} 
\end{eqnarray}
Here, $M_{L_{2,3}}, N_{L_{2,3}}$ and $B_{L_{2,3}}$ stand for edge-integrals of magnetic, non-magnetic and background signals, respectively. Note that in the difference of the PE-normalized spectra there are terms originating from non-magnetic and background signal. These can introduce a systematic error into $m_l/m_s$ evaluation.

Let's try to estimate a magnitude of this error according to sum rule, eq.~(\ref{eq:mlms}):
\begin{eqnarray*}
  \frac{m_l}{m_s} & = & \frac{2}{3} \frac{[M_{L_3} + a(N_{L_3}+B_{L_3})] + [-M_{L_2} + a(N_{L_2}+B_{L_2})]}{[M_{L_3} + a(N_{L_3}+B_{L_3})]-2[-M_{L_2} + a(N_{L_2}+B_{L_2})]} \\
                  & = & \frac{2}{3} \frac{M_{L_3}-M_{L_2}+a(N_{L_3}+N_{L_2}+B_{L_3}+B_{L_2})}{M_{L_3}+2M_{L_2}+a(N_{L_3}-2N_{L_2}+B_{L_3}-2B_{L_2})} \\
            & \approx & \left. \frac{m_l}{m_s}\right|_{net} + \frac{2}{3}\frac{a(N_{L_{2,3}}+B_{L_{2,3}})}{M_{L_3}+2M_{L_2}} \\
            & - & \frac{2}{3}a\frac{(M_{L_3}-M_{L_2})(B_{L_3}-2B_{L_2})}{(M_{L_3}+2M_{L_2})^2} \\
            & = & \left. \frac{m_l}{m_s}\right|_{net}\left( 1-a\frac{B_{L_3}-2B_{L_2}}{M_{L_3}+2M_{L_2}} \right) + \frac{2a}{3}\frac{N_{L_{2,3}}+B_{L_{2,3}}}{M_{L_3}+2M_{L_2}}
\end{eqnarray*}
where we assumed $N_{L_3} - 2N_{L_2} \ll B_{L_3}-2B_{L_2}$. 

The size of the last term can be estimated using Eqns.~(\ref{eq:amc}) and (\ref{eq:mnmax}):
\begin{equation}
  \frac{2a}{3}\frac{N_{L_{2,3}}+B_{L_{2,3}}}{M_{L_3}+2M_{L_2}} = \frac{0.004 |M_{max}|}{N_{L_3,max} (1+0.003)} \frac{N+B}{3M}
\end{equation}
where we assumed that $M_{L_3} \approx M_{L_2} \equiv M$. Assuming that the shapes of the magnetic peak and non-magnetic one do not vary much throughout the diffraction plane we get $M \propto M_\mathrm{max}$ and $N,B \propto N_{L_3,\mathrm{max}}$, i.e., the ratio $M_\mathrm{max}/N_{L_3,\mathrm{max}} \times (N+B)/M$ is a constant and its value for our dataset is approximately $2.24$, which gives an error estimation
\begin{equation}
  0.004 \times 2.24 \approx 0.01
\end{equation}
Therefore this term alone explains the difference between expected 0.019 and obtained 0.028 (MCR and DD) values of $m_l/m_s$ ratio.

The second error term is proportional to $m_l/m_s$. Ratio
\begin{equation}
  -a\frac{B_{L_3}-2B_{L_2}}{M_{L_3}+2M_{L_2}} = \frac{0.002|M_\mathrm{max}|}{N_{L_3,\mathrm{max}}} \frac{2B_{L_2}-B_{L_3}}{M}
\end{equation}
is again approximately constant throughout the diffraction plane, because $M \propto M_\mathrm{max}$ and $B_{L_{2,3}} \propto N_{L_3,\mathrm{max}}$. Its value is around $0.002 \times 5.7=0.012$, which after multiplication by $m_l/m_s$ gives a negligible correction 0.00023.






\begin{thebibliography}{99}
\bibitem{note1} For a perfect 3-beam orientation one can use a mirror-symmetry plane to extract the EMCD.
\end{thebibliography}
\end{document}